# The University of Washington Mobile Planetarium
## A Do-it-Yourself Guide

Phil Rosenfield, Justin Gaily, Oliver Fraser, John Wisniewski

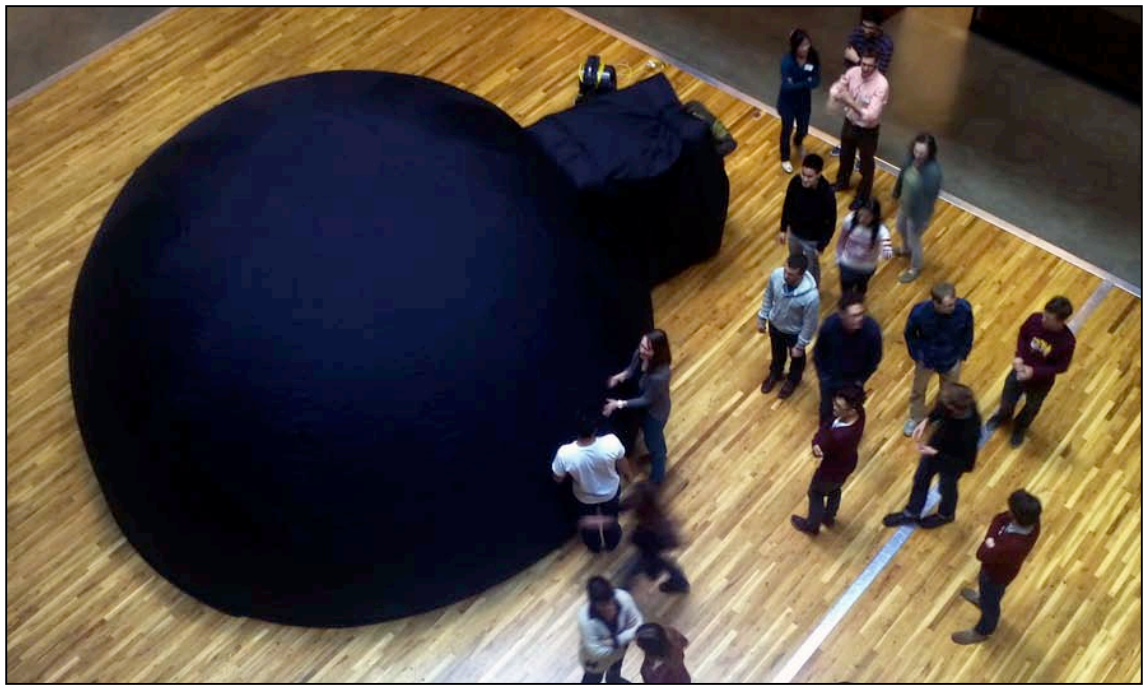

# Table of Contents

# Introduction

The UW Mobile Planetarium Project is a student driven effort to bring astronomy to high schools and the Seattle community. We designed and built an optics solution to project WorldWide Telescope in an inflatable planetarium from a laptop and off-the-shelf HD projector. In our first six months of operation, undergraduates at the UW gave planetarium shows to over 1500 people and 150 high school students created and presented their own astronomy projects in our dome, at their school. This document aims to share the technical aspects behind the project in order for others to replicate or adapt our model to their needs. This project was made possible thanks to a Hubble Space Telescope Education/Public Outreach Grant.

# Motivation

Digital planetariums are becoming a mainstay in astronomy education, as they allow the presenter to enhance their lessons with both incredible imagery that has become commonplace in the space age, as well as visualizations of astronomical systems from moons to galaxies. Free software, especially WorldWideTelescope (WWT[1]) has brought high quality, up-to-date astronomical imagery to the screens of anyone with an internet connection. Furthermore, WWT contains its own image warping software, making do-it-yourself planetariums with HD imagery within the reach of smaller budgets. In fact, the method we describe below cost roughly $18,000 in parts (all purchased new). Our costs would have been about $2,000 less had we already owned the laptop and projector. The largest cost is in the planetarium dome and first surface mirror ($12,000).

The mobile planetarium project grew from an existing planetarium outreach program. The graduate students in the University of Washington Astronomy department maintain a weekly outreach program where they organize and present free planetarium shows to any school or astronomy group that makes a reservation. In 2009, organizers noticed that in the three years prior, this outreach program was serving on average 1,000 students a year. However, in the same period, no Seattle Public High Schools made reservations, even though they are all located within 10 miles of the UW planetarium. We decided that a proactive solution to our lack of SPHS engagement was to bring our planetarium shows to the schools.

However, we quickly learned that with WWT software we no longer needed to lecture, and instead could "flip" the planetarium. That is, students could create and present their own planetarium shows. Our initial plan, to turn our planetarium outreach program into a road show, became simplistic and outdated in the face of new technology. We now engage students by helping them produce their own planetarium content, and bring our mobile planetarium to help them stage their astronomically themed presentations.

We now describe the technical decisions we made and provide advice we wish we had when starting this project from scratch.

# Before You Begin

### Access Existing Knowledge Bases
We strongly suggest becoming a member of the yahoo groups "full_dome" and "small_planetarium" and diving into their archives.

### Timeline
We planned on nine months of part time work to gather equipment, design and build optics housing, and test the optical alignment. We also allotted one quarter to offer a 1-credit seminar to train undergrads in setting up and operating the planetarium. Finally, we set up two meetings with a "pilot" classroom before launching into full operation.

### Our Budget
We received an HST/EPO grant for $40,000 to increase access to the UW planetarium and build a mobile planetarium. As stipulated, we were limited to spend no more than half of our funding on mobile planetarium equipment. In total, our planetarium cost $14,000 in parts which included purchasing a $1,500 laptop.

### Estimated Non-Equipment Costs

#### Personnel
This project was never full time for any person involved. The initial overhead is the highest concentration of labor, where the planetarium is built, and a team of undergraduates is trained in WWT software and the technical details of the planetarium. For this initial ramp up, we hired one graduate student and one undergraduate assistant.

The hired graduate student ordered and led the planetarium assembly as well as mentored the undergraduate assistant. The undergraduate assisted by leading the building and design of the optics housing as well as wrote lessons and trained prospective undergraduate presenters in a 1-credit seminar.
    Graduate Student: ~300 hours total, billed at 20% FTE for 9 months
    Undergraduate: ~240 hours total, billed hourly for 1 year.

With the planetarium in full swing, we have a team of approximately ten undergraduate volunteers who are capable of transporting the planetarium. Currently, one full time UW lecturer is in charge of the mobile planetarium, though we recommend keeping this position as an advisory role and funding 1-2 undergraduate assistants to manage the schedule and communication with the schools.

#### Insurance
Insurance is an important element to remember to include in a longer-term budget.

#### Transportation
Although we have transported our entire planetarium and a passenger inside a four-door sedan, typically we prefer to travel in groups of at least three people. In that case we rent a minivan. Depending on the range over which you expect to travel consider budgeting for rental vehicles and mileage costs.

# Equipment

## Essential Equipment

### Projection Type
We found that a fish-eye lens solution would be prohibitively expensive as a single purchase and in the event we need to find a replacement projector. With the idea that undergraduates would be transporting the equipment on their own and 30-35 high school students would be filing into our planetarium, we wanted a projection system that would sit on an edge of the dome, rather than the center. We opted to purchase two first-surface mirrors, one convex, and one flat, to project imagery on the dome.

### Inflatable dome, fan, and hemisphere mirror
The biggest equipment cost is the inflatable dome. The decision of which size dome and which company to use should be made with care. We will not reproduce the clearinghouse of knowledge and experience in the Yahoo groups, *small_planetarium* and *full_dome*. We made heavy use of their email archive as well as asked specific questions to the group at large. We list our main concerns and the solutions we came to with the help of the yahoo groups. Advice from the experiences of members in the yahoo groups positively mentioned Go-Dome, Digitalis, and Stargazer. In the end, we bought a standard sized Go-Dome through [eplanetarium.com](eplanetarium.com)[2], which came with an inflating fan and the hemisphere mirror.

#### *Dome size*
Concern: Transportable by car (by 2-3 undergraduate students), ability to fit class of (~30) high school students inside, and ability to fit within a classroom.

Solution: We limited our search for domes that were no more than 10' high.

New issues raised by our solution:
The horizon will be low, most students will need to sit on the ground, some chairs or perhaps two wheelchairs can be placed around the back and sides of a dome this size.

Why constrain to presentations in a classroom?
We could have purchased a taller dome and required that our set up would have to be in a gym, cafeteria, or theater (outside is not an option as any wind will cause the dome to lose its shape). We made our choice based on the following two issues:
1) We assumed there would be a no internet access outside of a classroom (in fact, we rarely have internet access in the schools)
2) We assumed it would be more difficult for a science class to take over the other locations and wanted our imprint on the school to be as small as possible (for example, it could be too loud to share a space). In fact we've found that librarians are often happy to share their space—but certainly it's been very helpful to have the option of using a classroom.

We recommend calling around to different schools at this point to see what options are available. In the end, we would have made the same decision on the dome size, and purchased the standard Go-Dome.

### *Dome entrance*
Concern: ADA compliant

Solution: We haven't found an excellent solution for inflatable domes. The best option seemed to be to purchase a standing dome (one that does not require constant inflation) that has an open entrance. In our research, these domes were well beyond our equipment budget. Advice from experiences of members in the yahoo groups positively mentioned Go-Dome, Digitalis, and Stargazer as ADA compliant options.

### *Dome material:*
Concern: Will the dome let in outside light? Is it safe to bring into schools? Has it been fire tested?

Solution: All the above domes are light tight. The three companies listed above all seemed to have dark domes and the necessary documentation.

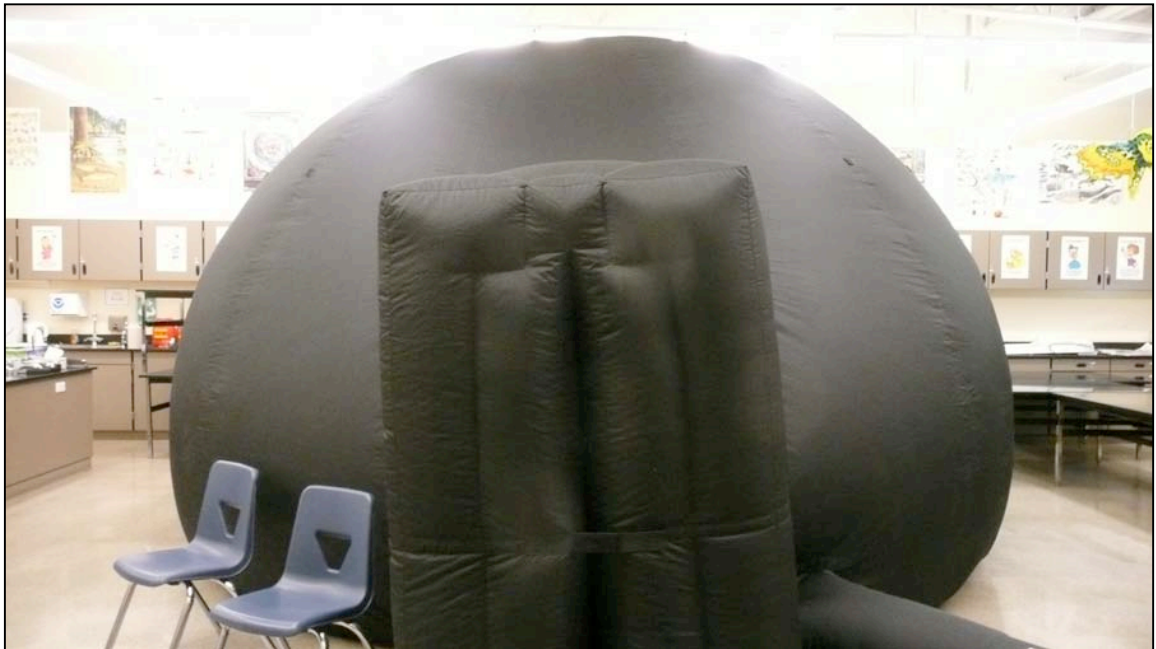

*The UW Mobile Planetarium dome in a high school science classroom (students are inside the dome)*

*Mirror costs*
Concern: First surface hemisphere mirrors are expensive, and seemed to be only produced in Australia. How can we limit the cost as we are based in the US?

Solution: First surface mirrors are a must. Coated mirrors produce blurry images as some of light from the projector is reflected by the interior surface of the coating back to the mirror, and travels to the dome at a new angle. This is only made worse if more than one mirror is used.

ePlanetarium ships a first surface hemisphere mirror for an additional cost with the Go-Dome. (They also have their own TSA-approved optics solution, which was beyond our budget, and may have limited our projector choice to a projector with a lens in its center.)

*Dome Fan*
Concern: Our only concern was the fan's portability in light of the rest of the equipment.

Solution: It's simple to purchase a small-wheeled attachment to the fan, or a luggage accessory that is a foldable two-wheeler. We haven't found the need to purchase them.

However, we quickly learned that fan speed, fan control, and fan noise are important factors. After observing that the fan needed to be turned up to high while people enter and exit the dome, we were concerned that we'd need a way to control the fan's speed from inside. However, we've realized that we actually always have someone on the outside to assist with crowd control, and that they can use the fan's speed to communicate to the presenter. We typically turn up the fan when it's time to wrap up the show.

In a small room, a large fan can create a lot of background noise. We suggest looking closely into the specifics of the dome fan to make sure it fits your needs.

**Projector**
Concern: Small budget, good for high dark-light contrast (stars and nebulosity), easy portability, small replacement costs.

Solution: We purchased an off-the-shelf, 1920x1080p (16x9) HD, high lumen projector. Projector Central[3] is a powerhouse of information when it comes to choosing projectors. We limited our search to 1920x1080p (16x9) HD projectors under $1000. We find that high lumen projectors are better suited for the mobile planetarium purposes. In large planetariums we make use of the dark adaption so with a low lumen projector, the eye can better pick out details like constellations after seeing a bright image. In our small planetarium, however, the line of sight to the image is never more than 15ft, and usually around 10ft. We cannot depend on the dark adaption of students' eyes, for example, after flashing an image of the Hubble Space Telescope's mosaic of the Crab Nebula spread on the entire dome. Finally, we pay no attention to the quoted contrast ratio, since dynamic irises and other technologies make the quantity non-uniformly defined from projector to projector.
We are aware of planetariums that have chosen to use LED projectors, which have the advantage of not needing to replace bulbs, and would consider that in the future.

**Laptop**

Concern: HD video card, large hard drive space, Windows PC or Mac running Windows on a dual boot or as parallels (for WWT).

Solution: Any laptop with a video card capable of extending an HD display and dedicated hard drive space for WWT to cache imagery is fine. Look for one with a backlit keyboard so the presenter can type in the dark (a USB powered reading light would be an affordable work around to a backlit keyboard). From our experience (and not industry comparison) we have been happy with a near top of the line NVIDIA GeForce video card. In simpler terms, the laptop should have a built in (mini) DV or HDMI output. For lower quality imagery, VGA can be used, but is not recommended.

**Optics assembly**
Concern: Mainly durability, size, and cost. We wanted to limit the handling of the first surface mirror(s).

Solution: To save money, we designed and built our own. Full details of our solution are posted on our website[4] and available from the authors.

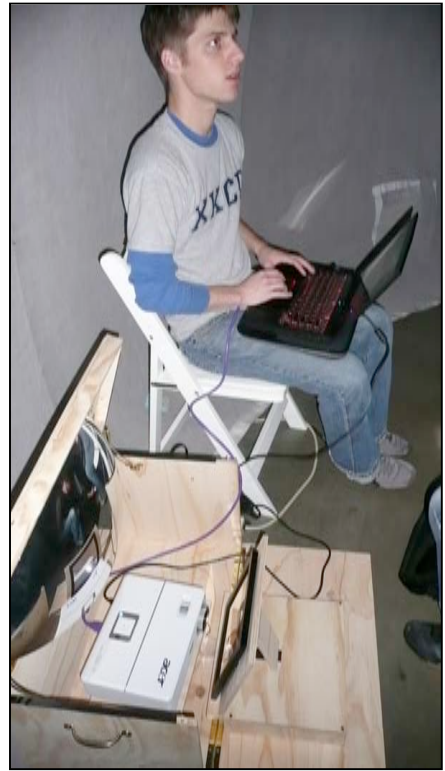

*Justin Gailey tests his optics box in the mobile planetarium. The DYI guide to building the above optics box is available on our website[4].*

**Essential Accessories**

**Cables and extension cords**

*Power*
The laptop, the projector, lights, and perhaps other accessories such as speakers and PA systems, require power. It is often against fire code to connect a power-strip to an extension cord, so in our case it was important to purchase a single unit. Our originally 30 foot outdoor extension cord with three outlets has been replaced with a power strip with eight outlets and a 25 foot cord. We typically plug the fan in directly to a room outlet, alongside with our extension cord.

*Display*
Not all HD projectors come with DVI or HMDI cables, and some laptops need a cable to convert HDMI to mini DV. Using only the VGA cable that comes with an HD projector is like buying a sports car and never taking it out of 2$^{nd}$ gear.

## Non-Essential Equipment

### Secondary mirror
We recommend a secondary flat first surface mirror. It allows the projector to be safely placed underneath the hemisphere mirror, and thus takes up less physical space in the planetarium, meaning more places for humans, and a smaller chance of being bumped and jostled. However, it adds more variables to the alignment.

### Equipment cases

#### *Dome*
The inflatable Go-Dome from ePlanetarium came with two bags. It was shipped in one bag that was rubberized and contained a canvas bag with a zipper. We found it was difficult to replace the dome after inflation in either bag, as well as transport something so heavy without wheels. Instead, we purchased a rolling equipment bag made for hockey goalies. At the time of our purchasing, goalie bags prices ranged from $80-$120. Ours is large enough to fit extra smaller equipment and not require expert dome repackaging. With some extra budget, we would have had a logo option!

#### *Mirrors*
The hemisphere mirror is the most delicate and difficult to replace item, as there is no repair for scratches. We built its housing as part of the box we would transport it in, to avoid the number of times it would be handled (taking something in and out of a box for every show seemed to ask for scratches)

The secondary mirror is less than 12" diameter and kept in a picture frame, which is covered with cardboard and sealed with rubber bands, so that nothing touches the mirror surface. We keep it in our laptop bag.

#### *Laptop*
A simple laptop backpack is enough to hold the laptop, lots of cables, a mouse, an Xbox controller, non-essential accessories, and any paperwork (such as the fire retardancy certificate and our contact information). A backpack is nice as it keeps a hand free to carry other equipment.

#### *Projector*
Most off the shelf projectors come with a carrying bag. With the amount of travel— in and out of cars and schools while carrying other equipment—we decided to purchase a heavy-duty pelican case for the projector. We included the cost in the projector budget.

### Lighting
It is always nicer to have people enter a planetarium with more lighting than the brightest lights the planetarium will get. We use a simple solution of rope lights around the edge of the dome, and bought a small switch so the presenter has easy access to turn the "house lights" on and off. We also use a battery powered camping lantern to provide light during setup and takedown.

### Non-Essential Accessories

#### Audio and Public Address (PA) Equipment
WWT can play prerecorded tours with audio, which requires some sort of amplified speaker system. We've found that speakers placed outside the dome work great, as do higher quality computer speakers placed near the presenter. We now use a wireless microphone worn by the presenter, another wireless microphone that can be passed around, and a small PA system to mix the microphone sounds and the laptop. This equipment is an investment of about $1,000.

#### Tickets
We hand out tickets for particular show times when we are presenting at school science nights, which typically have us doing many short shows in a row. The tickets ensure that people know when to show up for their show, and they allow us to manage the number of people at each show. We print 30 tickets at each show time on pull-apart business card stock, which allows us to easily print more when we need them. With this system people know when to return, we manage the crowds, and we collect the tickets to use again afterward.

#### Laptop cooling pad
For longer shows, sitting with a hot laptop can get uncomfortable. We recommend a cooling pad that is long enough to hold a mouse.

#### Seating
A carpet, a bunch of carpet squares, small benches, or folding seats that give back support would probably be a welcomed accessory for those seated on the floor.

### Non-Essential Equipment and Accessories for WWT

#### Internet access
WWT caches imagery from servers around the world. We purchased a 30ft long Ethernet cable as a back up for internet access. Another possibility is using a wireless card in the laptop. We found that neither the Ethernet cord nor a wireless card is essential. If weak or no internet is available, this is a useful site[5], as is the WWT documentation housed on their site.

#### Mouse/Xbox controller
WWT can be operated with a USB Xbox controller as well as a mouse. Some of our presenters are happy rolling a standard mouse around on top of the keyboard, although a 3D mouse is a natural fit here.

#### MIDI Controller
WWT also has the ability to attach a MIDI controller and customize its key bindings. This is an excellent solution to the keyboard/mouse control, but takes some work to set up (we have yet to implement it).

## Initial Assembly

### Optics Box Construction
Justin Gaily, who designed and lead the building of our Optics box, has written a separate DYI guide, posted on our [website](#) and available from the authors.

### Testing and Alignment
With the optics box ready, it was great to have high ceiling rooms to align and test our system, as well as train undergraduates. The UW Dance and Theater departments graciously provided these spaces. WWT makes warping very easy in several scenarios, including a 16x9 mirror dome (see WWT documentation for details). The rest of the setup involved adjusting the components of the optics box: positions of the projector and angles of the mirrors until the entire dome is filled with light. It is helpful to project a grid during this process.

## "Flipping" the Planetarium
In our experience, one measure of a successful education/public outreach project is how well it can be adapted to the specific needs of the target market. We wrote our grant with the simple idea to bring our successful planetarium program directly to the Seattle schools and community, but we have discovered that students can create their own tours of the universe in our planetarium.

Our model is to support teachers during a "planetarium presentations unit" lasting one or two weeks. The unit begins with small groups of students choosing a topic in astronomy and creating a storyboard for a short (3-5 minute) presentation using imagery from WWT. If the teacher isn't trained in WWT, we make an initial visit to the classroom to demonstrate WWT tour creation and check in with each student group. After this visit, students work together to create WWT tours. Finally, our team returns with the mobile planetarium and the students present their work to their peers.

Students create a story as they research their topic, and then practice their communications skills to present it. On the presentation day, everyone gets to see their tour projected inside the dome!

We have found that students have no problem creating tours that show well in the planetarium (as long as they avoid projecting text). We advise them to consider that only the middle third of their computer's screen will be in front of them when they are inside the dome and there is no reading from scripts inside the dome, so they can either record a voiceover or memorize what they want to say.

## Creating Tours

General information on creating tours is available here.

We use the following script to quickly show a class some tour making ideas, as well as demonstrating the capability of WWT:

1. Quick tour of WWT
    a. solar system mode
    b. turn constellations off
    c. double click on a planet to go there
    d. sky mode!
    e. demonstrate search
2. Tour creation
    a. Pull down Tour menu and create a new tour
    b. create new slide
    c. change view
    d. set end position
    e. create the next slide right away for a smooth transition
    f. save!
3. Demonstrate the tour!
    a. Play the tour while reciting lots of numerical information
    b. Then play it while revealing interesting facts
    c. Which is more interesting?
4. Tour creation notes
    a. Avoid zooming in so close to your subject that it extends into the top or sides of the screen, because in the mobile planetarium the top of the screen projects onto the ceiling, and the sides project to either side of the audience, and your subject will appear inside-out!
    b. Text usually is hard to read in the mobile planetarium.

## Contact

Email our team: uw.mobile.planetarium@gmail.com
Find us online: http://www.astro.washington.edu/groups/outreach/mplanetarium/

---

[1] www.worldwidetelescope.com
[2] www.eplanetarium.com
[3] www.projectorcentral.com
[4] http://www.astro.washington.edu/groups/outreach/mplanetarium/about.html
[5] http://blogs.msdn.com/b/astrojonathans_hack-a-day_blog/archive/2013/04/15/running-worldwide-telescope-quot-off-the-grid-quot-offline-cache-management.aspx